\documentclass[prl,twocolumn,amsmath,amssymb,superscriptaddress,
showpacs]{revtex4}
\usepackage{graphics}
\usepackage{epsfig}
\usepackage{bbm}
\usepackage[latin1]{inputen c} 

\newcommand{\be}{\begin{equation}}
\newcommand{\ee}{\end{equation}}
\newcommand{\eea}{\end{eqnarray}}

\newcommand{\va}[1]{\ensuremath{(\Delta#1)^2}}
\newcommand{\vat}[1]{\ensuremath{(\tilde{\Delta}#1)^2}}

\newcommand{\ex}[1]{\ensuremath{\left\langle{#1}\right\rangle}}
\newcommand{\exs}[1]{\ensuremath{\langle{#1}\rangle}}

\newcommand{\qed}{\ensuremath{\hfill \Box}}

\newcommand{\ketbra}[1]{\ensuremath{| #1 \rangle \langle #1 |}}

\newcommand{\ket}[1]{\ensuremath{|#1\rangle}}

\newcommand{\bra}[1]{\ensuremath{\langle#1|}}

\newcommand{\kommentar}[1]{}
\newcommand{\trace}{{\rm Tr}}

\newcommand{\coll}{A}
\newcommand{\collB}{a}

\begin{document}
\title{
Spin squeezing inequalities for arbitrary spin}
\date{\today}
\begin{abstract}
We determine the complete set of generalized spin squeezing
inequalities, given in terms of the collective angular momentum components,
for particles with an arbitrary spin. 
They can be used for the experimental detection of entanglement in
an ensemble  in which the particles cannot be individually
addressed. We also present a large set of
criteria involving collective observables different from the angular momentum coordinates.
We show that some of the inequalities
can be used to detect $k$-particle entanglement and bound entanglement.
\end{abstract}

\author{Giuseppe Vitagliano}
\affiliation{Department of Theoretical Physics, The University of the Basque Country,
P.O. Box 644, E-48080 Bilbao, Spain}

\author{Philipp Hyllus}
\affiliation{Department of Theoretical Physics, The University of the Basque Country,
P.O. Box 644, E-48080 Bilbao, Spain}

\author{I\~nigo L. Egusquiza}
\affiliation{Department of Theoretical Physics, The University of the Basque Country,
P.O. Box 644, E-48080 Bilbao, Spain}

\author{G\'eza T\'oth}
\affiliation{Department of Theoretical Physics, The University of the Basque Country,
P.O. Box 644, E-48080 Bilbao, Spain}
\affiliation{IKERBASQUE, Basque Foundation for Science, E-48011 Bilbao, Spain}
\affiliation{Research Institute for Solid State Physics and Optics, Hungarian
Academy of Sciences, P.O. Box 49, H-1525 Budapest, Hungary}


\maketitle



With an interest towards fundamental questions in quantum physics, as well as applications,
larger and larger entangled quantum systems have been realized with photons, trapped ions and cold atoms \cite{review}. 
Quantum entanglement
can be used as a resource for certain quantum
information processing tasks \cite{review}, and it is also
necessary for a wide range of interferometric schemes
to achieve the maximum sensitivity in metrology \cite{PS09}.
Hence, the verification of the presence of entanglement
is a crucial but exceedingly challenging task, especially in 
an ensemble of many, say $10^6-10^{12},$ 
particles. In such systems, typically the particles are not accessible individually  and
only collective operators can be measured.
A ubiquitous entanglement criterion in this context is
the spin squeezing inequality
 \cite{SD01}
\begin{eqnarray}
\frac{\va{J_x}}{\exs{J_y}^2+\exs{J_z}^2}\ge \frac{1}{N},
\label{motherofallspinsqueezinginequalities}
\end{eqnarray}
where $N$ is the number of spin-$\frac{1}{2}$ particles, 
$J_l:=\sum_{n=1}^N j_{l}^{(n)}$ for $l=x,y,z$ are the
collective angular momentum components and $j_{l}^{(n)}$ are the
single spin angular momentum components acting on the $n^{th}$ particle. If a state violates
Eq.~(\ref{motherofallspinsqueezinginequalities}),
then it is entangled (i.e., not fully separable \cite{W89}).
Such spin squeezed states \cite{K93}
have been created in numerous experiments with
cold atoms and trapped ions
\cite{HS99,review}, and
can be used, for instance, in atomic clocks
to achieve a precision higher than the shot noise limit \cite{K93}.

Recently, after several generalized spin squeezing inequalities (SSIs) for the
detection of entanglement appeared in the literature
\cite{GT04,KC05,GT06} and were used experimentally
\cite{spexp}, a complete set of such entanglement conditions has been
presented in Ref.~\cite{TK07}. However, all of the above mentioned
conditions are for spin-$1/2$ particles (qubits), and so far the literature on
systems of particles with $j>\frac{1}{2}$ is
limited to a small number of conditions, specialized
for certain quantum states or particles with a low dimension
\cite{DC02,GT04,MK08}. At this point the question arises: Could
one obtain a complete set of inequalities for $j>\tfrac{1}{2}$?
Such conditions would be very relevant from the practical point of view since in
most of the experiments the physical spin of the particles is larger than
$\tfrac{1}{2}$ and the spin-$\tfrac{1}{2}$ subsystems are
created artificially. Thus, knowing the full set of entanglement criteria
for $j>\tfrac{1}{2}$, many experiments for realizing large scale entanglement
could be technologically less
demanding, and fundamentally new experiments could also be carried out. The solution is not simple:
Known methods for detecting entanglement for spin-$\frac{1}{2}$ particles by spin-squeezing cannot straightforwardly
be generalized to higher spins. For example, for $j > \frac{1}{2},$
Eq.~(\ref{motherofallspinsqueezinginequalities})
can also be violated without entanglement between the spin-$j$ particles
\cite{F08}.

In this Letter, we present the complete set of optimal spin squeezing
inequalities for the collective angular momentum coordinates
for a system of $N$ particles with spin $j.$ 
We also show how existing entanglement conditions for spin-$\frac{1}{2}$ particles
can be transformed into entanglement conditions for spin-$j$ particles
with $j>\frac{1}{2}$ (i.e., qudits with a dimension $d=2j+1$).
Finally, we present a large set of entanglement conditions
for qudit systems that involve operators different from the
angular momentum coordinates, and investigate in detail one of the conditions.

{\em Definitions.}  The basic idea for the qudit case is that besides $j_l,$ other single-qudit quantities can also be measured. Let us consider particles with  $d$ internal states.   $a_k$ for $k=1,2,...,M$ will denote single-particle operators with the property  
$\trace(a_k a_l)=C\delta_{kl},$ where $C$ is a constant. As we will show later, the $a_k$ operators can be, for instance, the SU(d) generators for a $d$ dimensional system. 
Moreover, for obtaining our generalized spin squeezing inequalities, we will need the upper bound $K$ for the inequality  $\sum_{k=1}^{M}\exs{\collB_{k}^{(n)}}^{2}\le K.$ 

The $N$-qudit collective operators used in our criteria will be denoted by $\coll_{k}=\sum_{n}\collB_{k}^{(n)}.$  In the qubit case, the SSIs were  developed based on the first and second moments and variances of the such collective operators \cite{TK07}. For $j>1/2,$
we define the modified second moment \begin{equation}\langle \tilde{\coll}_{k}^{2}\rangle:=\langle\coll_{k}^{2}\rangle-\langle\sum_{n}(\collB_{k}^{(n)})^{2}\rangle=\sum_{m\ne n}\langle\collB_{k}^{(n)}\collB_{k}^{(m)}\rangle\label{mod1}\end{equation} and the modified variance
\begin{equation}\vat{{\coll}_{k}}:=\va{\coll_{k}}-\exs{\sum_{n}(\collB_{k}^{(n)})^{2}}.\label{mod2}\end{equation}
In the following, the quantities Eq.~(\ref{mod1}) and Eq.~(\ref{mod2}) will be used instead of second moments and variances because
otherwise it is not possible to obtain tight inequalities for separable states \cite{MK08}.

{\em SSIs for qudits.} First, we present a general inequality from which the entanglement conditions for the different operator sets
can be obtained. \\
{\bf Observation 1.}---For separable states, i.e., for states that can be written as a mixture of product states \cite{W89},
\begin{equation}
(N-1)\sum_{k\in I}\vat{{\coll}_{k}}-\sum_{k\notin I}\exs{\tilde{\coll}_{k}^{2}}\ge-N(N-1)K \label{spinsqsud}
\end{equation}
holds, where each index set $I \subseteq\{1,2,...,M\}$ defines one of the $2^M$ inequalities.
Note that $I=\emptyset$ and $I=\{1,2,...,M\}$ are among the possibilities. The proof can be found in the Appendix.
It is remarkable that the bound on the right-hand side of Eq.~(\ref{spinsqsud}) is tight, independent of $I,$
and independent of the particular choice of the $a_k$ operators except for the value of $K.$

Equation~(\ref{spinsqsud}) is the basis for the entanglement conditions we present in Obs. 2 and 4. 

{\bf Observation 2.}---Optimal spin squeezing inequalities for qudits. For fully
separable states of spin-$j$ particles, all the following inequalities are fulfilled
\begin{subequations}
\begin{eqnarray}
\exs{J_x^2}+\exs{J_y^2}+\exs{J_z^2} &\le& Nj(Nj+1),
\label{theorem1a}
\\
\va{J_x}+\va{J_y}+\va{J_z} &\ge& Nj, \label{Jxyzineq_singlet}
\\
\exs{\tilde{J}_k^2}+\exs{\tilde{J}_l^2}-N(N-1)j^2 &\le&
(N-1)(\tilde{\Delta} J_m)^2, \label{Jxyzineq_spsq2}
\\
(N-1)\left[(\tilde{\Delta} J_k)^2+(\tilde{\Delta} J_l)^2\right]
&\ge& \exs{\tilde{J}_m^2}-N(N-1)j^2, \label{Jxyzineq_spsq3}
\;\;\;\;\;\;
\end{eqnarray} \label{Jxyzineq}\end{subequations}
where $k,l,m$ take all possible permutations of $x,y,z.$
Violation of any of the inequalities (\ref{Jxyzineq}) implies
entanglement. 
The inequalities (\ref{Jxyzineq}) are a full set
for large $N$ in the sense that it is not possible to add a new entanglement condition detecting other states based on $\exs{J_k}$ and $\exs{\tilde{J}_k^2}.$\\
{\it Proof.}---We applied Observation 1 with $\{a_k\}=\{j_x,j_y,j_z\},$ $K=j^2$ and used $j_x^2+j_y^2+j_z^2=j(j+1)\openone$ \cite{angular,SUPP}. 
For $j=\frac{1}{2},$ the inequalities (\ref{Jxyzineq}) are identical to the optimal SSIs for qubits \cite{TK07}. For this case, the completeness  has already been shown \cite{TK07}. That is, for all values
of $\exs{J_k}$ and $\ex{\tilde{J}_k^2}$ that fulfill Eqs.~(\ref{Jxyzineq})
there is a corresponding separable state in the large $N$
limit. Direct calculation shows that if a separable quantum state
$
\varrho_{{\rm sep},\frac{1}{2}}=\sum_m p_m \rho_m^{(1)} \otimes \rho_m^{(2)} \otimes...\otimes\rho_m^{(N)},
$
where $\rho_m^{(n)}$ are single-qubit pure states,
saturates one of the inequalities Eqs.~(\ref{Jxyzineq}) for $j=\frac{1}{2},$ then the state
$
\varrho_{{\rm sep},j}=\sum_m p_m \omega_m^{(1)} \otimes \omega_m^{(2)} \otimes...\otimes\omega_m^{(N)},
$
saturates the same inequality of Eqs.~(\ref{Jxyzineq}) for spin-$j$ particles.
Here, $\omega_m^{(n)}$ are single-qudit pure-state density matrices
such that $\trace(\rho_m^{(n)}\sigma_l)j=\trace(\omega_m^{(n)}j_l).$
For instance, if the first state is $\ket{+\frac{1}{2}}_x,$ then the second one is
$\ket{+j}_x.$ Thus the proof of completeness of Ref.~\cite{TK07} can be extended
to prove the completeness of the criteria Eqs.~(\ref{Jxyzineq}). \qed

Eq.~(\ref{theorem1a}) is valid for
all quantum states.
States maximally violating Eq.~(\ref{Jxyzineq_singlet}) are angular momentum singlets,
while for Eq.~(\ref{Jxyzineq_spsq2}), for even $N,$ they are symmetric Dicke states of the form
$\binom{N/2}{N}^{-\frac{1}{2}}\sum_k \mathcal{P}_k (\ket{+j}^{\otimes N/2}\otimes \ket{-j}^{\otimes N/2}),$ where $\mathcal{P}_k$
denotes all different permutations \cite{TOBEPUBLISHED}. 

It is also possible to obtain entanglement conditions for spin-$j$ particles
from criteria for qubit systems.\\
{\bf Observation 3.}---Let us consider an inequality valid for $N$-qubit separable states
of the form
\begin{equation}
f(\{\exs{J_l}\}_{l=x,y,z},\{\exs{\tilde{J}_l^2}\}_{l=x,y,z})\ge {\rm const.},\label{fff}
\end{equation}
where $f$ is a concave function of its variables.
All of the generalized SSIs in the literature have this form.
Then, the entanglement condition Eq.~(\ref{fff})
can be transformed to a criterion for a system of $N$ spin-$j$ particles
by the substitution
\begin{eqnarray}
\exs{J_l} &\rightarrow& \tfrac{1}{2j}\exs{J_l}, \;\;\;\;\;\;
\exs{\tilde{J}_l^2}\rightarrow
\tfrac{1}{4j^2}(\exs{\tilde{J}_l^2}).
\end{eqnarray}
{\it Proof.}---
Let us consider product states of $N$ spin-$j$ particles of the form $\varrho_{j}=\otimes_n \varrho_j^{(n)},$ and define the quantities $r_l^{(n)}=\exs{j_l^{(n)}}/j.$ Then, the first and second moments can be rewritten as  $\exs{J_l}=j\sum_n r_l^{(n)}$ and
$\exs{\tilde{J}_l^2}=j^2\sum_{m\ne n} r_l^{(n)} r_l^{(m)}.$
The only constraint for the physically allowed values for $r_l^{(n)}$ is
 $\vert \vec{r}^{(n)}\vert \le 1$ for all $j.$ Hence, for an arbitrary function $f,$
\begin{eqnarray}
&&\min_{\varrho_j}f(\{\tfrac{1}{2j}\exs{J_l}_{\varrho_{j}}\}_{l=x,y,z},\{\tfrac{1}{4j^2}\exs{\tilde{J}_l^2}_{\varrho_{j}}\}_{l=x,y,z})\nonumber\\
&&\;\;\;\;\;\;\;\;\;=\min_{\varrho_{1/2}}f(\{\exs{J_l}_{\varrho_{1/2}}\}_{l=x,y,z},\{\exs{\tilde{J}_l^2}_{\varrho_{1/2}}\}_{l=x,y,z}).\nonumber
\end{eqnarray}
If $f$ is a concave function of its variables then
we have the same minimum for separable states. \qed

Using Observation 3, for instance, the standard spin-squeezing inequality
Eq.~(\ref{motherofallspinsqueezinginequalities}) from
Ref.~\cite{SD01} becomes
\begin{eqnarray}
\frac{\va{{J}_x}}{\exs{J_y}^2+\exs{J_z}^2}+\frac{\sum_n (j^2-\exs{(j_x^{(n)})^2})}{\exs{J_y}^2+\exs{J_z}^2}\ge \frac{1}{N}. \label{motherofallspinsqueezinginequalities2}
\end{eqnarray}
Equation~(\ref{motherofallspinsqueezinginequalities2}) is violated
only if there is entanglement between the spin-$j$ particles.
Because of the second, nonnegative term on the left-hand side of 
Eq.~(\ref{motherofallspinsqueezinginequalities2}), for $j>\frac{1}{2}$
there are states that violate Eq.~(\ref{motherofallspinsqueezinginequalities}),
but do not violate Eq.~(\ref{motherofallspinsqueezinginequalities2}).
Remarkably, it can be proven that Eq.~(\ref{Jxyzineq_spsq2}) is strictly stronger than Eq.~(\ref{motherofallspinsqueezinginequalities2}) \cite{TOBEPUBLISHED}.

The last application of Obs. 1 is the following.\\
{\bf Observation 4.}---For a system of $d$-dimensional particles, we can define collective operators based on the SU(d) generators $\{g_k\}_{k=1}^M$ with $M=d^2-1$ as $G_k=\sum_{n=1}^N g_k^{(n)}.$
The SSIs for $G_k$ have the general form
\begin{eqnarray}
&&(N-1)\sum_{k\in I}\vat{{G}_{k}}-\sum_{k\notin I}\exs{\tilde{G}_{k}^{2}}\ge\nonumber\\
&&\;\;\;\;\;\;\;\;\;\;\;\;\;\;\;\;\;\;\;\;\;\;\;\;\;\;\;\;\;\;\;\;\;\;-2N(N-1)\frac{(d-1)}{d}.\label{SU3general}
\end{eqnarray}
For instance, for the $d=3$ case, the SU(d) generators can be the Gell-Mann matrices \cite{GM62}.\\
{\it Proof.}---We used Observation 1 with $C=2$ and 
$K=2(1-\frac{1}{d})$ 
\cite{LOO,SUPP}. \qed

Observation 4 presents an abundance of inequalities. Here, we will analyze in detail Eq.~(\ref{SU3general}) for $I  =\{1,2,...,M\}.$  
Using $\sum_k g_k^2=2(d+1)(1-\frac{1}{d})\openone$ \cite{SUPP}, 
Eq.~(\ref{SU3general})  for  this case can be rewritten  as \begin{equation}
\sum_{k=1}^{d^{2}-1}\va{G_{k}}\ge 2N(d-1).\label{SU3singlet}
\end{equation}
Equation~(\ref{SU3singlet}) is maximally violated by many-body SU(d) singlets. 
Such states appear often in statistical physics of spin systems and condensed matter physics \cite{GR07}. They are invariant under operations of the type $U^{\otimes N}$ \cite{W89},
which can be exploited in differential magnetometry \cite{TM10}, encoding quantum
information in decoherence free subspaces and sending information independent from the reference frame direction \cite{LC98}.

{\em Noise tolerance of Eq.~(\ref{SU3singlet}).} First, we will ask how efficiently Eq.~(\ref{SU3singlet}) can be used for entanglement
detection. Let us consider SU(d) singlet states (i.e., states with $\exs{G_k^2}=0$) mixed with white noise as
$\varrho_{{\rm noisy}}=(1-p_{{\rm noise}})\varrho_{{\rm singlet}}+p_{{\rm noise}}\frac{1}{d^{N}}\openone.$
Direct calculation shows that such a state is detected as entangled if
$p_{{\rm noise}}<\frac{d}{d+1}.$ 
Thus, the noise tolerance in detecting SU(d) singlets is increasing with $d$.
Note that Eq.~(\ref{Jxyzineq_singlet}) detects a noisy state as entangled for an analogous
situation if $p_{{\rm noise}}<\frac{2}{d+1}.$

{\em Eq.~(\ref{SU3singlet}) detects $k$-particle entanglement.} The criteria presented so far detect any type of non-separability. It would be important to
find similar criteria that detect higher forms of entanglement, that is, $k$-entanglement. This type of
strong entanglement, rather than simple non-separability,
 is needed, for instance, to achieve maximal precision in many interferometric tasks \cite{fisher_kentanglement}.
 A pure state is said to possess $k$-entanglement if it cannot be written as a tensor product $\otimes_n \ket{\psi_n}$ such that
 each $\ket{\psi_n}$ is a state of at most $k-1$ qubits. A mixed state is $k$-entangled if it cannot be obtained mixing states that are at most $k-1$ entangled \cite{GT05}. Otherwise the state is called $(k-1)$-producible.

While Eq.~(\ref{SU3singlet}) can be maximally violated by two-producible states for $j=\frac{1}{2}$ \cite{TM10}, it is not the case for
$j>\frac{1}{2}.$
For the SU(d) case, a $d$-particle entangled state is needed to violate Eq.~(\ref{SU3singlet}) maximally \cite{SUPP}. Thus, the amount of violation of Eq.~(\ref{SU3singlet}) can be used to detect $k$-entanglement.\\
{\bf Observation 5.}---For two-producible states the following bound holds
\begin{equation}
\sum_{k=1}^{d^{2}-1}\va{G_{k}}\ge \bigg\{
                                       \begin{array}{ll}
                                         2N(d-2) &\text{ for even }N, \\
                                         2N(d-2)+2 &\text{ for odd }N. \\
                                       \end{array}
\label{twoprod}
\end{equation}
The violation of Eq.~(\ref{twoprod}) signals $3$-particle entanglement.
Note that for large $d$ the bound in Eq.~(\ref{twoprod})
is very close to the bound for separable states in Eq.~(\ref{SU3singlet}). 
The proof  can be found in the Appendix. 

{\em Eq.~(\ref{SU3singlet}) detects bound entanglement.} In Ref.~\cite{TK07}, it has already been shown the optimal SSIs for the $j=\frac{1}{2}$ case
can detect bound entanglement \cite{bound}, i.e., entangled states with a positive partial transpose (PPT,  \cite{ppt}), in the thermal states of common spin models. We find numerically that the criterion Eq.~(\ref{SU3singlet}) detects bound entanglement in the thermal state of
several Hamiltonians, such as for example $H=\sum_k G_k^2,$ even for $j>\frac{1}{2}$ 
\cite{TOBEPUBLISHED}.

{\em Symmetric states.} Next, it is important to ask how our entanglement criteria behave for symmetric states, as such states naturally appear in many systems such as Bose-Einstein condensates of two-state atoms.\\
{\bf Observation 6.}---
(i) Symmetric states can violate Eq.~(\ref{spinsqsud})  for some $I$ only if 
$\varrho_{{\rm av}2}^{T1}\ngeqslant0,$ where T1 denotes the
partial transposition \cite{ppt} and the average two-qudit density matrix is defined as
$\varrho_{{\rm {\rm av}2}}=\frac{1}{N(N-1)}\sum_{m\ne n} \varrho_{mn}.$
(ii) For symmetric states, if  $a_k$ are the SU(d) generators $g_k,$ 
Eq.~(\ref{spinsqsud}) is equivalent to
\begin{equation}
\sum_{k\in I}N\vat{{G}_{k}}+\exs{G_{k}}^{2}\ge0.\label{symsq}
\end{equation}
For this case, Eq.~(\ref{symsq}) is violated for at least one $I$     and some choice of the collective operators if and only if
$\varrho_{{\rm av}2}^{T1}\ngeqslant0.$
For the proof, see the Appendix. 

{\em Implementation.} The angular momentum coordinates $J_k$ and their variances can be measured in cold atoms by coupling the atomic spin to a light field, and then measuring the light \cite{HS99}. The collective spin can be rotated by magnetic fields.
Measuring the operators $\sum_n (j_k^{(n)})^2$ can be realized by rotating the spin by a magnetic field, and then measuring the populations of the $j_z$ eigenstates. In some cold atomic systems, such operators might also be measured directly, as in such systems in the Hamiltonian a $s_z (j_k^{(n)})^2$ term appears, where $\vec{s}$ is the photonic pseudospin \cite{KM10}.
For the SU(d) generators, the $G_k$ operators can be measured in a similar manner, however,
SU(2) rotations realized with a magnetic field are not sufficient.
For larger spins, it is advantageous to choose the $g_k$ operators to be
 $(\ket{k}\bra{l}+\ket{l}\bra{k})/\sqrt{2},$ $i(\ket{k}\bra{l}-\ket{l}\bra{k})/\sqrt{2}$ and
  $\ketbra{k}$ \cite{LOO2}.
The corresponding collective operators can all be measured based an SU(2) rotation within
a two-dimensional subspace and a population measurement of at most
two quantum states.

In summary, we have presented a complete set of generalized SSIs
for detecting entanglement in an ensemble of qudits based on knowing only
$\exs{J_k}$ and $\exs{\tilde{J}^2_k}$ for $k=x,y,z.$ We extended our approach to collective observables
based on the SU(d) generators. We showed that some of the inequalities can be used to detect $k$-entanglement
and bound entanglement. Finally, we discussed the experimental 
implementation of the criteria. 
\begin{acknowledgments}
We thank O. G\"uhne and Z. Kurucz for discussions. We thank the ERC StG GEDENT\-QOPT,
the MICINN
(Project No. FIS2009-12773-C02-02),
the Basque Government (Project No. IT4720-10), and
the National Research Fund of Hungary OTKA (Contract No. K83858).
\end{acknowledgments}

{\it Appendix.}---{\it Proof of Observation 1.} We consider product states of the form
$\ket{\Phi} = \otimes_n \ket{\phi_n}.$
For such states, we have
$\ex{\tilde{\coll}_{k}^2}_{\Phi} =\exs{A_k}^2-\sum_{n}\exs{\collB_{k}^{(n)}}^2.$
Hence, the left-hand side of Eq.~(\ref{spinsqsud}) equals
$-\sum_{n}(N-1)\sum_{k\in I}\exs{\collB_{k}^{(n)}}^{2}
-\sum_{k\notin I}\left(\exs{\coll_{k}}^{2}-
\sum_{n}\exs{\collB_{k}^{(n)}}^{2}\right)\ge
-\sum_{n}(N-1)\sum_{k=1}^{M}\exs{\collB_{k}^{(n)}}^{2}\ge-N(N-1)K.$
We used that $
\exs{\coll_{k}}^{2}
\le
N\sum_{n}\exs{\collB_{k}^{(n)}}^{2}$ \cite{TK07}. 
\qed

{\it Proof of Observation 5.}  We will find a lower bound on the left-hand side of Eq.~(\ref{twoprod})
for $N=2.$ Let us consider first antisymmetric states. We will use that $\sum_k \langle  G_k^2 \rangle = \sum_k \langle  g_k^2\otimes\openone \rangle +
\sum_k \langle  \openone \otimes g_k^2 \rangle + 2 \sum_{k} \langle  g_k \otimes g_k \rangle.$
Then, we need that  $\sum_{k} g_k \otimes g_k =2F-\frac{2}{d}\openone$
where $F$ is the flip operator \cite{LOO,SUPP}. Hence, $\sum_k \langle  G_k^2 \rangle = 4(d+1)(1-\frac{2}{d}).$
For the nonlinear part, we have that $\sum_k \langle g_k \rangle^2_{\varrho_{\rm red}} = 2 {\rm Tr}(\varrho_{\rm red}^2) - \frac{2}{d}$ \cite{LOO,SUPP}, and using the Cauchy-Schwarz inequality for $\sum_k \exs{g_k\otimes\openone}\exs{\openone\otimes g_k},$
we obtain a bound
$\sum_k \langle  G_k \rangle^2\le 4-\frac{8}{d}.$ Here we used that for antisymmetric states, for the reduced single-qudit state
$\trace(\varrho_{\rm red}^2)\le \frac{1}{2}$  \cite{SC01}.
This leads to Eq.~(\ref{twoprod}) for antisymmetric states.
For symmetric states the bound on the left-hand side of
Eq.~(\ref{twoprod}) can be obtained similarly and it is larger. Finally, since the equation is invariant under the permutation of qudits,
the variances give the same value for $\varrho$ as for $\frac{1}{2}(\varrho+F\varrho F)\equiv P_a \varrho P_a + P_s \varrho P_s,$
where $P_s$ and $P_a$ are the projectors to the symmetric and antisymmetric subspaces, respectively.
Thus, it is sufficient to consider mixtures of symmetric and antisymmetric states.
The bound for the product of such two-qudit states and of single-qudit states for the left-hand side of Eq.~(\ref{twoprod}) can be obtained using $[\Delta (a\otimes\openone+\openone\otimes a)]^2_{\psi_1\otimes\psi_2} =\va{a}_{\psi_1}+\va{a}_{\psi_2}.$ Because of the concavity of the variance,
the bound is the same for mixed $2$-producible states. \qed

{\it Proof of Observation 6.}  Equation~(\ref{spinsqsud}) can be rewritten as $\sum_{k\in I}N\vat{{\coll}_{k}}+\exs{\coll_{k}}^{2}\ge\sum_{k=1}^{M}\exs{\tilde{A}_{k}^{2}}-N(N-1)K,$
which can be reexpressed as
$\sum_{k\in I} N\left(\exs{\collB_{k}\otimes \collB_{k}}_{\varrho_{{{\rm av}2}}}-\exs{\collB_{k}\otimes\openone}_{\varrho_{{{\rm av}2}}}^{2}\right)\ge
\sum_k \exs{a_k \otimes a_k}_{\varrho_{{{\rm av}2}}}-K.$ From Eq.~(\ref{spinsqsud}) for  $I=\emptyset$ it follows that $\sum_k \exs{a_k \otimes a_k}_{\varrho_{{{\rm av}2}}}=\frac{1}{N(N-1)}\sum_k\exs{\tilde{A}_k^2}\le K,$ while the equality holds for symmetric states for the SU(d) generators $g_k$ \cite{SUPP}.
We also need that a density matrix of a two-qudit symmetric state has a positive partial transpose if and only if
 $\exs{O\otimes O}-\exs{O\otimes\openone}^2\ge 0$ for every $O$ \cite{TG09}.
Hence the statement of Observation 6
follows.
For qubits, we obtain the results of Ref.~\cite{KC05}. \qed

\eject

\renewcommand{\thefigure}{S\arabic{figure}}
\renewcommand{\thetable}{S\arabic{table}}
\renewcommand{\theequation}{S\arabic{equation}}
\setcounter{figure}{0}
\setcounter{table}{0}
\setcounter{equation}{0}
\newcommand{\loo}{{\lambda}}

\subsection{Supplementary Material}

The supplement contains some derivations to help to understand the details of
the proofs of the main text. It summarizes well-known facts
about the quantum theory of angular momentum and that of SU(d) generators.
More details will be presented elsewhere [S1].

\textbf{\emph{Angular momentum operators.}} Next, we summarize the fundamental equations
 for angular momentum operators [S2]. For particle with spin-$j$ we have
\begin{equation}
(j_x^2+j_y^2+j_z^2)=j(j+1)\openone.\label{xyz}
\end{equation}
Since the angular momentum operators have identical
      spectra, it follows from Eq.~(\ref{xyz}) that we can write
\begin{equation}
{\rm Tr}(j_x^2)=\frac{1}{3}j(j+1)(2j+1). \label{trjx2}
\end{equation}
Based on Eq.~(\ref{trjx2}),
 we get the constant for the orthogonality relation
\begin{equation}
{\rm Tr}(j_k j_l)=\delta_{kl}\frac{1}{3}j(j+1)(2j+1).
\end{equation}

For the sum of the squares of expectation values we have
\begin{equation}
\sum_{k=x,y,z}\exs{j_k}^2\le j^2.\label{j2}
\end{equation}
For $j=\frac{1}{2},$ for all pure states the equality holds for
Eq.~(\ref{j2}).

Finally,
\begin{equation}
\sum_{l=x,y,z}\langle (j_l \otimes \openone+\openone\otimes j_l)^2\rangle \le 2j(2j+1).
\end{equation}
Hence, using Eq.~(\ref{xyz}) we obtain
\begin{equation}
2j(j+1)+2\sum_{l=x,y,z}\langle j_l \otimes j_l\rangle \le 2j(2j+1).
\end{equation}
Thus, we arrive at the inequality
\begin{equation}
\sum_{l=x,y,z}\langle j_l \otimes j_l\rangle \le j^2.
\end{equation}

\textbf{\emph{Local orthogonal observables.}} Here we summarize the results of Ref.~[S3] for Local Orthogonal Observables (LOOs, [S4]). For a system of dimension $d,$ these are $d^2$ observables $\loo_k$ such that
\begin{equation}
\trace(\loo_k\loo_l)=\delta_{kl}.
\end{equation}
For a quantum state $\varrho,$ LOOs have the following properties
\begin{eqnarray}
\sum_{k=1}^{d^2} (\loo_k)^2 &=& d \openone,\\
\sum_{k=1}^{d^2} \langle \loo_k \rangle ^2 &=& {\rm Tr}(\varrho^2)\le 1.
\end{eqnarray}
Moreover, based on Ref.~[S5] we know that
\begin{equation}
\sum_{k=1}^{d^2} \loo_k \otimes \loo_k = F,
\end{equation}
where $F$ is the flip operator exchanging two qudits.

\textbf{\emph{SU(d) generators.}} Next, we will use the results known for local orthogonal
observables for SU(d) generators. For a system of dimension $d,$ there are $d^2-1$ traceless SU(d) generators $g_k$ with the property
\begin{equation}
\trace(g_kg_l)=2\delta_{kl}.
\end{equation}
Thus, from SU(d) generators $g_k$ we can obtain LOOS using
\begin{equation}
\loo_k=\frac{1}{\sqrt{2}}g_k
\end{equation}
for $k=1,2,...,d^2-1,$ and $\loo_{d^2}=\frac{1}{\sqrt{d}}\openone.$

After a derivation similar to that of Ref.~[S3], we arrive at
\begin{equation}
\sum_{k=1}^{d^2-1} (g_k)^2 = 2\frac{d^2-1}{d} \openone,\label{square}
\end{equation}
\begin{equation}
\sum_{k=1}^{d^2-1} \langle g_k \rangle ^2 = 2\left({\rm Tr}(\varrho^2)-\frac{1}{d}\right)\le 2\left(1-\frac{1}{d}\right),
\end{equation}
\begin{equation}
\sum_{k=1}^{d^2-1} g_k \otimes g_k = 2\left(F-\frac{1}{d}\openone\right).\label{ggg}
\end{equation}
Based on Eq.~(\ref{ggg}), for bipartite symmetric states we have
\begin{equation}
\langle \sum_{k=1}^{d^2-1} g_k \otimes g_k \rangle= 2\left(+1-\frac{1}{d}\right),
\end{equation}
while for antisymmetric states we have
\begin{equation}
\langle \sum_{k=1}^{d^2-1} g_k \otimes g_k \rangle= 2\left(-1-\frac{1}{d}\right).
\end{equation}
It is important to stress that the inequalities presented are valid
for all SU(d) generators, not only for Gell-Mann matrices.

\textbf{\emph{Equations for the collective operators based on SU(d) generators.}}

Here we present some fundamental relations for the collective operators $G_k.$
First of all, the length of the vector $\vec{G}=\{\exs{G_k}\}_{k=1}^{d^2-1}$ is maximal for
a state of the form $\ket{\Psi}^{\otimes N}.$ This can be seen as for such states
$\vec{G}=N\vec{g}$ where $\vec{g}=\{\exs{g_k}_{\Psi}\}_{k=1}^{d^2-1},$
and knowing that for pure states $\vert\vec{g}\vert$ is maximal.

For the sum of the squares of $G_k$ we obtain
\begin{eqnarray}
\sum_k (G_k)^2&=&\sum_k \sum_n  (g_k^{(n)})^2+\sum_k \sum_{n\ne m}  g_k^{(m)}g_k^{(n)}\nonumber\\
              &=&2N\frac{d^2-1}{d} \openone+ \sum_{n\ne m}  2\left(F_{mn}-\frac{\openone}{d}\right).\nonumber\\ \label{sumg2}
\end{eqnarray}
Here we used Eq.~(\ref{square}) and Eq.~(\ref{ggg}).
Based on Eq.~(\ref{sumg2}) and using $\exs{F_{mn}}\ge -1$, we can write
\begin{equation}
\sum_k \exs{(G_k)^2}\ge\frac{2N}{d} (d+1)(d-N).\label{asym}
\end{equation}
Note that the bound on the right-hand side of Eq.~(\ref{asym}) cannot be zero if $N<d.$
For $N=d,$ the sum $\sum_k \exs{(G_k)^2}$ is zero for the totally antisymmetric state for which
$\exs{F_{mn}}=-1$ for all $m, n.$

Next, we will show that
\begin{equation}
\sum_k \langle G_k^2 \rangle=0\;\;\;\Leftrightarrow\;\;\;\sum_k (\Delta G_k)^2=0. \label{leftright}
\end{equation}
In order to prove that, one has to notice that $\sum_k (\Delta G_k)^2=0$ implies
$\sum_k (\Delta G_k')^2=0$ for any set of SU(d) generators $G_k'$ [S6]. This also implies
$\va{B}=0$ for all traceless observables $B.$ For every traceless $D$ one can find traceless
$B_1$ and $B_2$ such that $[B_1,B_2]=iD $ [S7] and hence $\va{B_1}+\va{B_2}\ge \vert \exs{D}\vert.$ Hence, $\sum_k \exs{(G_k)^2}=0$ implies
$\exs{D}=0$ for all traceless observables $D$ [S1].

As a consequence of Eq.~(\ref{asym}) and Eq.~(\ref{leftright}), for $N<d$ we have $\sum_k (\Delta G_k)^2>0.$
Hence, for $d$-dimensional systems states with less than $d$ particles cannot have $\sum_k (\Delta G_k)^2=0.$

Moreover, for symmetric states we have $\exs{F_{mn}}=+1$ for all $m, n,$ and based on Eq.~(\ref{sumg2}) we obtain
\begin{equation}
\sum_k \exs{(G_k)^2}= \frac{2N}{d} (d-1)(d+N),
\end{equation}
which is the maximal value for $\sum_k \exs{(G_k)^2}.$
Similarly, for symmetric states,
\begin{eqnarray}
\sum_k \exs{(\tilde{G}_k)^2}&=&\sum_{k}\exs{(G_{k})^{2}}-\langle\sum_{k}\sum_{n}(g_{k}^{(n)})^{2}\rangle
\end{eqnarray}
is also maximal.

Naturally, these statements are also true for the angular momentum operators for the $j=\frac{1}{2}$ case,
as these operators, apart from a constant factor, are SU(2) generators.

On the other hand, for the angular momentum operators for $j>\frac{1}{2}$ these statements are not true. In particular, $\exs{\sum_k (J_k)^2}$ is not maximal for every symmetric state.

\vskip1cm

{\bf References}

\begin{enumerate}

\item[{[}S1{]}] G. Vitagliano, P. Hyllus, I.L. Egusquiza, and G. T\'oth, in preparation.

\item[{[}S2{]}] D. M. Brink and G. R. Satchler, {\it Angular momentum}, (Oxford University Press, USA, third edition, 1994).

\item[{[}S3{]}] O. G\"uhne {\it et al.},
Phys. Rev. A {\bf 74}, 010301 (2006).

\item[{[}S4{]}] S. Yu and N.-L. Liu, Phys. Rev. Lett. {\bf 95,} 150504 (2005).

\item[{[}S5{]}] G. T\'oth and O. G\"uhne, Phys. Rev. Lett. {\bf 102}, 170503 (2009).

\item[{[}S6{]}] Note that $\sum_k \va{G_k}=\trace(\gamma),$ where the covariance matrix is defined as $\gamma_{kl}=\frac{1}{2}(\langle \Delta G_k \Delta G_l \rangle+\langle \Delta G_l \Delta G_k \rangle).$
    $\trace(\gamma)$ is independent of the particular choice of the $G_k$ matrices.

\item[{[}S7{]}] This is true because the group generated by $G_k$ is a simple group. See also L. O'Raifeartaigh, {\it Group structure of gauge theories} (Cambridge University Press, New York, 1986).

\vfill

\end{enumerate}


\begin{thebibliography}{99}
\bibitem{LOO2} For these $d^2$ operators we have $K=1$ 
\cite{SUPP}.
\end{thebibliography}
\end{document}